# HARMONI at ELT: Design & Test on the current status of Low Order Wavefront Subsystem Pick Off Arm Engineering Model (LPOA-EM)

Iago Funes Vecino*[a], Guillermo Mercant Rubio[a], Alonso Álvarez Urueña[a], Laura García Moreno[a], Gonzalo José Carracedo Carballal[a], Irene M. Ferro Rodríguez[a], Heribert Argelaguet Vilaseca[a], Javier Piqueras López[a], on behalf of the HARMONI Consortium.

[a]Centro de Astrobiología (CAB) CSIC-INTA, Ctra. Ajalvir, Km 4.5, 28850 Torrejon de Ardoz, Madrid, Spain

*ifunes@cab.inta-csic.es

## ABSTRACT

HARMONI is the first-light, adaptive-optics-assisted, near-infrared integral field spectrograph for the ELT, covering 800–2450 nm with resolving powers from 3000 to 7000. The Low Order Wavefront Sensor subsystem (LOWFS) provides field stabilisation by measuring the motion of the instrument focal plane relative to the telescope wavefront sensors. Following a rescope process in 2025, LOWFS has been updated to operate three LOWFS Pick-Off Arms (LPOA) simultaneously within a shared volume. We present first results from the LPOA Engineering Model (EM) characterisation campaign on two rotary joints — shoulder and elbow — including measurements of angular resolution, discrete step tracking validated by independent metrology, rotation axis wobble, and radial and axial runout of the bearing assembly.

## 1. INTRODUCTION

HARMONI is the first light, adaptive optics assisted, near-IR integral field spectrograph for the ELT [1]. It covers a spectral range from 800nm to 2450nm with resolving powers from 3000 to 7000 and spatial sampling of 25mas and 6mas. It can operate in two adaptive optics modes - SCAO (including a high contrast capability), MCAO and NoAO. The project is resuming its final design phase after a rescope design phase in 2025.

The Low Order Wavefront Sensor subsystem (LOWFS) measures the motion of the instrument's focal plane relative to the telescope wavefront sensors, providing field stabilisation across all observing modes. The rescope has introduced significant changes to the LOWFS architecture, most notably the requirement to operate three LOWFS Pick-Off Arms (LPOA) simultaneously within the same dedicated volume. Each LPOA carries a relay mirror that must be positioned over a wide angular range with sub-arcsecond accuracy, placing stringent demands on angular resolution, positional repeatability, and mechanical quality of the bearing assemblies.

To assess whether the selected design meets these demands prior to the final design phase, an Engineering Model (EM) of the LPOA has been assembled and tested. The EM is based on the pre-rescope mechanical design, adapted to the new three-arm operating context [2] [3]. The test campaign has focused specifically on those performance aspects that can only be verified on a physical model: the encoder angular resolution at the micro-step level, the fidelity of discrete step movements as confirmed by independent external metrology, and the mechanical quality of the rotary joints in terms of axis wobble and axial runout. This paper presents the first results of that campaign, covering the shoulder and elbow joints of the LPOA EM.

# 2. METHODS

## 2.1 LPOA Engineering Model

The LPOA EM comprises two rotary joints — called shoulder (LPOAS) and elbow (LPOAE) — each driven by a permanent magnet synchronous motor and equipped with an optical incremental angular encoder. The two joints together define the pointing direction of the relay mirror over a range of ±90° at the shoulder. The test campaign was conducted independently on each joint.

## 2.2 Instrumentation

In addition to the onboard encoder, three external instruments were deployed as independent position references:

- **Autocollimator** — measures the angular orientation of a flat mirror in two orthogonal axes (µrad), used exclusively for rotation axis wobble characterization.
- **Interferometer** — measures linear displacement (µm), used as an independent position validator for discrete step movements.
- **Capacitive sensors**— measure the gap to the shaft surface (µm), used for radial and axial runout characterization of the bearing assembly.

No hardware trigger was available between instruments; motion onset was detected independently in each signal and the records were aligned in post-processing.

## 2.3 Tests performed

- **Angular resolution.** The motor was commanded a staircase of discrete micro-steps of increasing size, each held at a fixed dwell for several repetitions. The minimum step reliably distinguished from the encoder noise floor was determined.
- **Step tracking.** The motor executed a sequence of discrete steps of fixed amplitude, repeated over multiple cycles in both directions. The encoder response was cross-validated against the interferometer, converting measurements to a common linear unit via the known arm length.
- **Rotation axis wobble**. A flat mirror was mounted perpendicular to the shoulder shaft. The autocollimator recorded the mirror orientation over multiple sweep passes, characterising how the rotation axis tilts as a function of arm angle.
- **Radial and axial runout.** The capacitive sensors recorded the gap to the lateral shaft surface (radial) and to the end face (axial) during rotation.

# 3. RESULTS

The following sections present the results obtained for each measurement type. Where both joints were tested, results are reported side by side to allow direct comparison of the shoulder and elbow performance. It should be noted that these are the initial results of an ongoing campaign; some measurements are currently available for one joint only and will be extended to both joints in future test runs.

### 3.1 Short term resolution

The characterisation of the short-term resolution began by determining the smallest commanded step detectable by the encoder, establishing the noise floor of each joint independently. A step is considered resolved when its measured amplitude exceeds three times the standard deviation of the signal in at least 95% of trials. The shoulder joint achieves a noise-limited resolution of $4.34\times10^{-6}$ degrees (≈ 0.08 µrad), with a measured noise floor of $1.45\times10^{-6}$ degrees, as shown in Fig. 1. The elbow joint yields a resolution of $1.11\times10^{-4}$ degrees (≈ 1.9 µrad), with a noise floor of $3.69\times10^{-5}$ degrees, approximately 25 times higher than the shoulder (Fig. 2). In both cases, the minimum step empirically resolved in the measurement sequence is consistent with the noise-limited estimate. On the elbow joint, the encoder response was additionally validated against an interferometer, which measured individual steps of 0.44 µm with an in-window noise of 0.22 µm, in close agreement with the encoder-derived displacement.

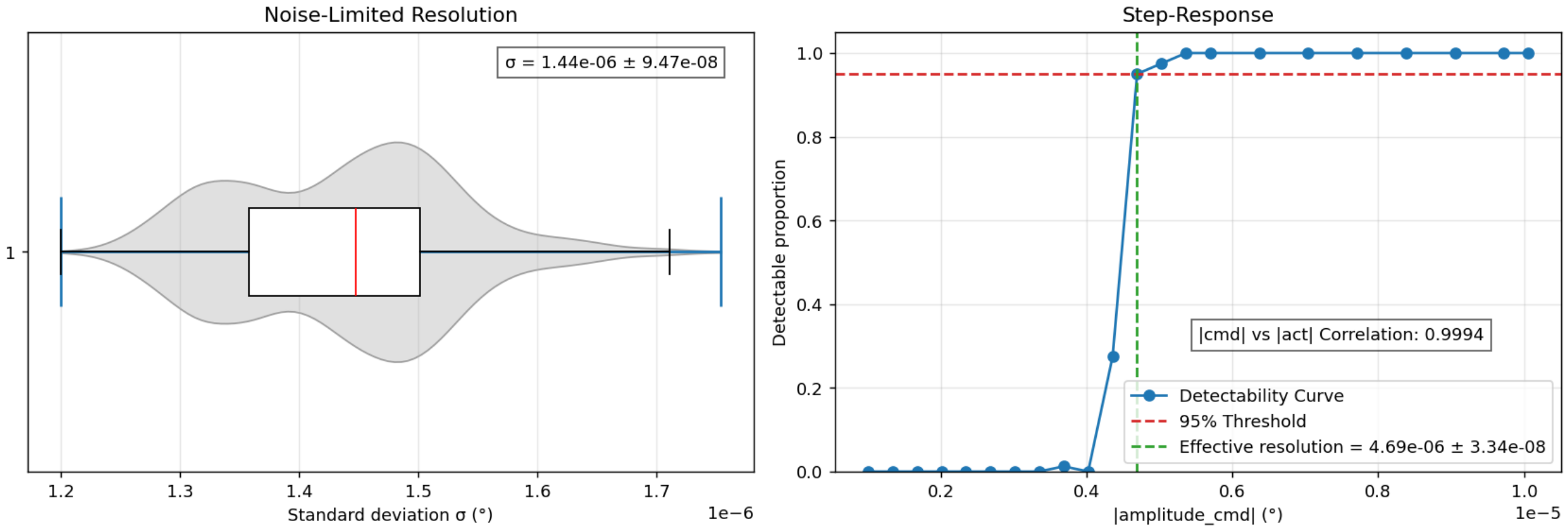

**Figure 1.** LPOAS Step resolution

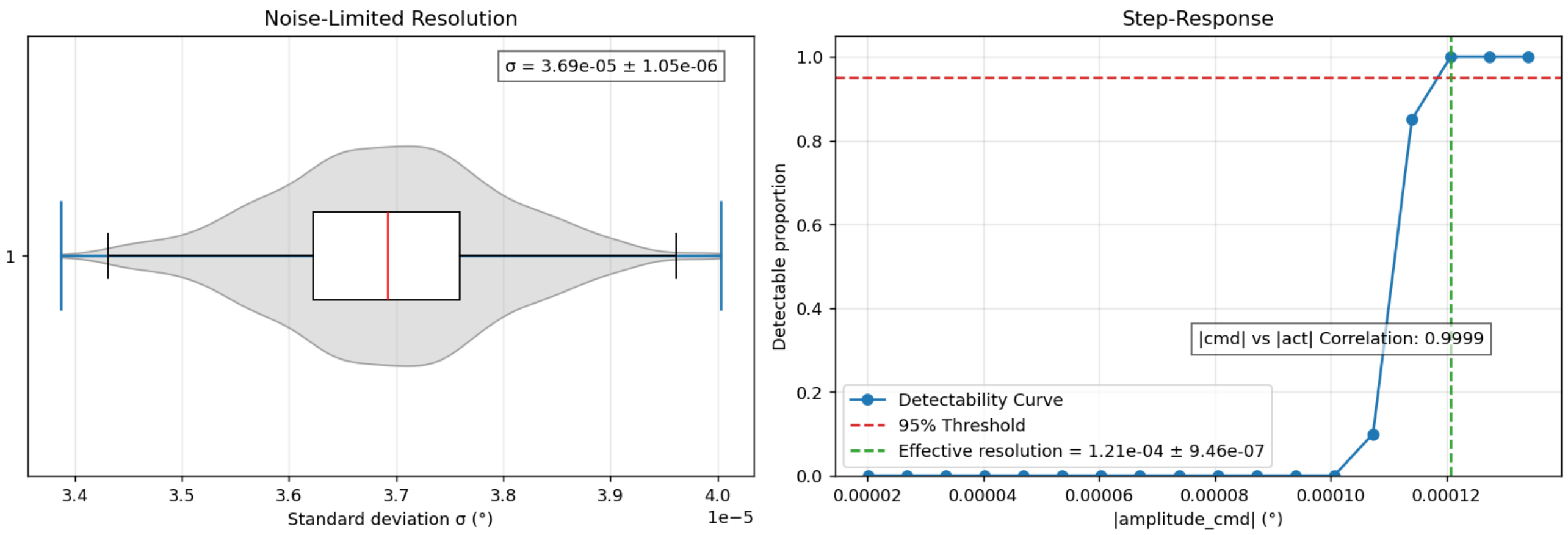

**Figure 2.** LPOAE Step resolution

### 3.2 Wobble

The motion of the shoulder rotation axis relative to its design ideal was investigated using an autocollimator and a folding mirror mounted perpendicular to the shaft (Figure 3).

The autocollimator signal in both measurement axes exhibits a smooth sinusoidal as a function of the joint angle (Fig 4), indicating that the axis tilt is dominated by a single once-per-sweep harmonic. The fitted amplitudes are 1278 µrad (Y-axis) and 1322 µrad (Z-axis), yielding a combined 2D wobble amplitude of 1839 µrad, as shown in Figure 4. Once the dominant harmonic is removed, the higher-order residual is 17.0 µrad peak-to-peak and 4.9 µrad RMS, representing less than 1.3% of the total wobble amplitude. The repeatability of the wobble across successive passes confirms that it is a stable characteristic of the bearing's geometry.

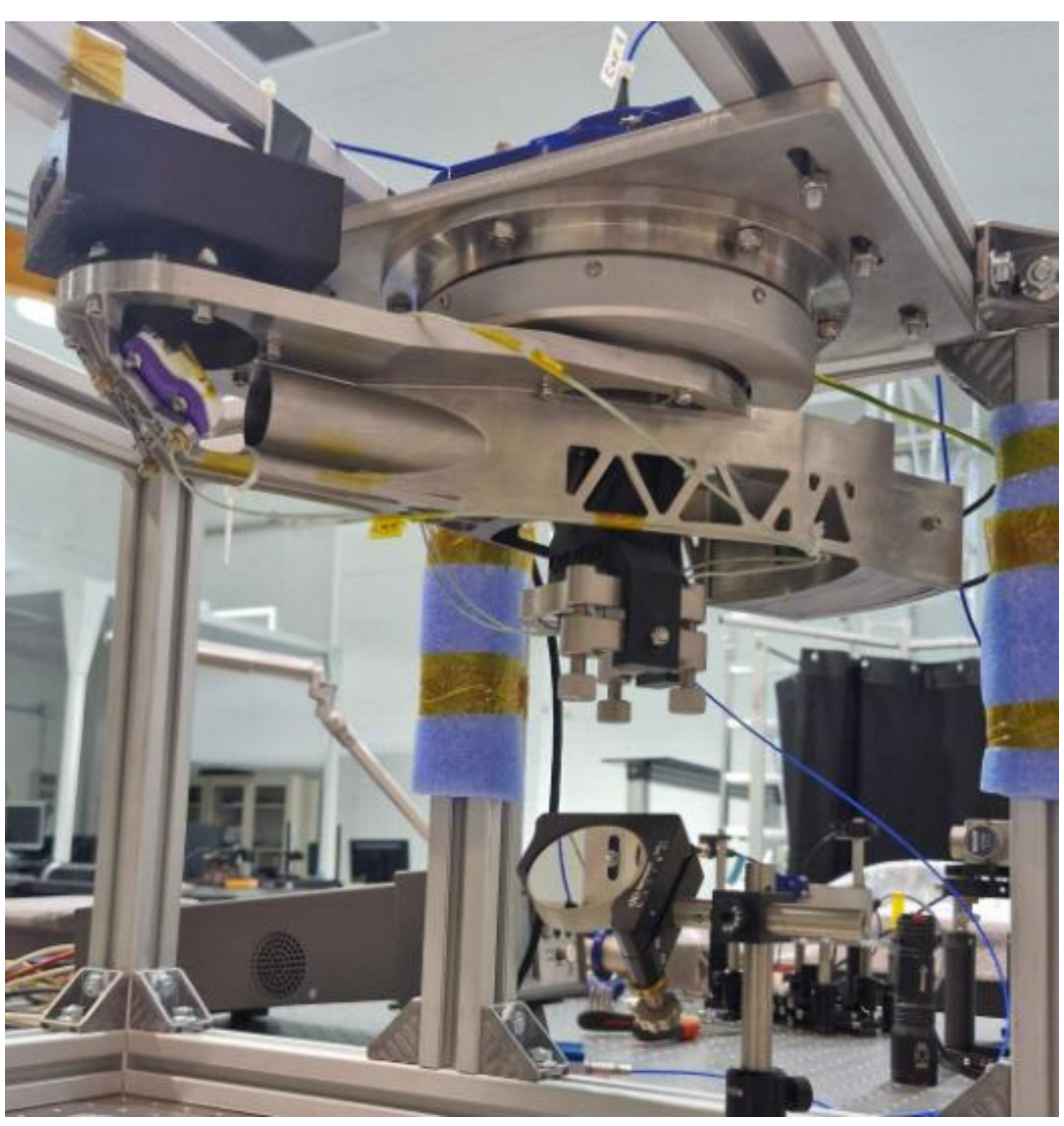

**Figure 3.** Experimental setup for wobble measurement — autocollimator and folding mirror mounted perpendicular to the shaft.

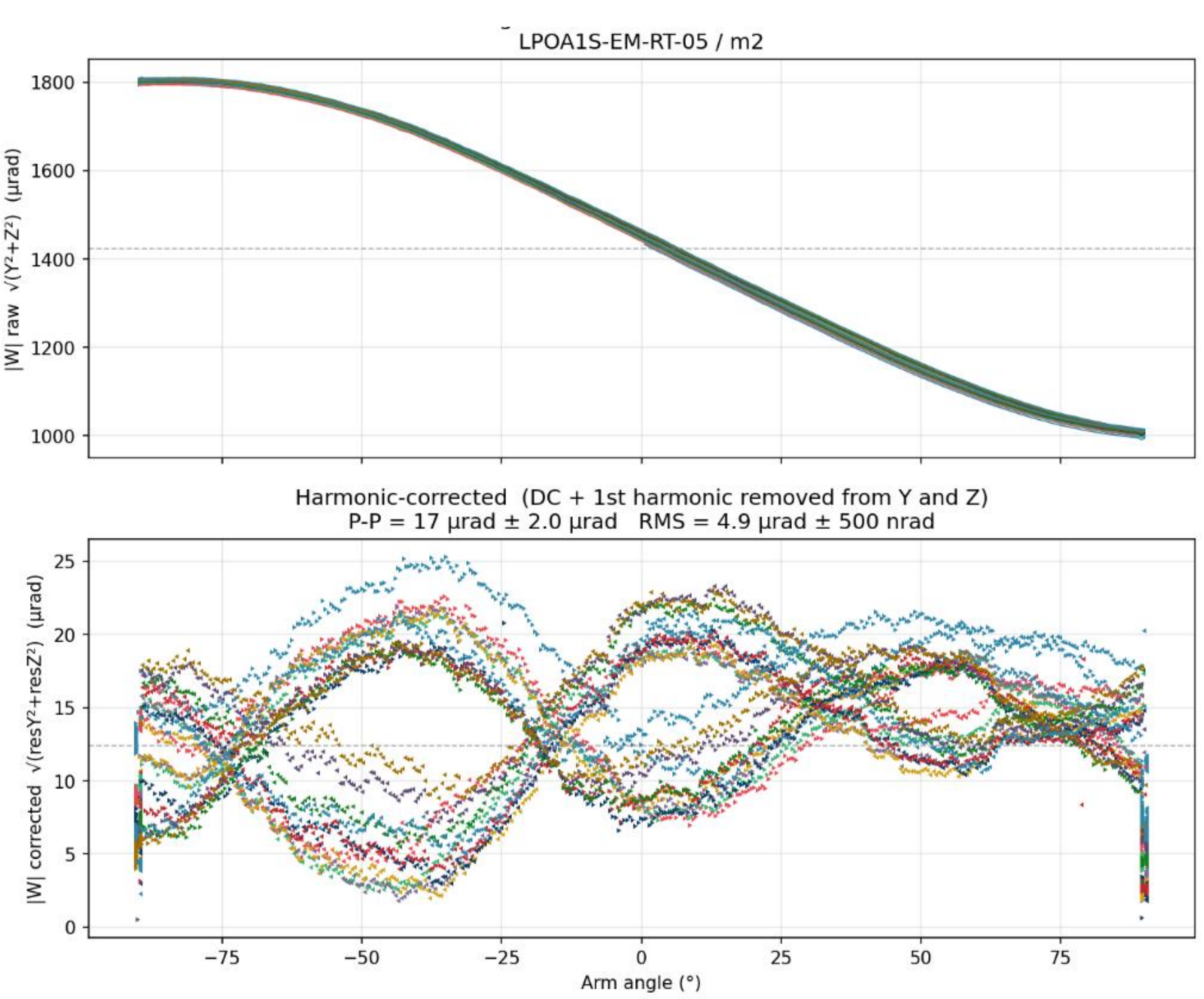

**Figure 4.** LPOAS Wobble magnitude

### 3.3 Radial runout

Two capacitive sensors placed 90° apart recorded the variation in distance to the lateral surface of the shaft during rotation, capturing both eccentricity components independently at X-Y plane (Fig 5).

The raw signal contains a dominant first-harmonic contribution due to the offset between the rotation axis and the sensor measurement axis; once this eccentricity term is removed, the residual reflects the true roundness deviation of the shaft surface. The eccentricity values, of the order of tens of µm on both joints, are a characteristic of the test setup rather than a defect of the shaft. The corrected runout profiles are shown in Fig. 6 for the shoulder and Fig. 7 for the elbow. The shoulder yields a residual roundness deviation of 1.2 µm peak-to-peak and 350 nm RMS. The elbow presents a higher deviation, with 14 µm peak-to-peak and 4.2 µm RMS, with successive revolutions overlapping closely in both cases, confirming that the deviations are repeatable surface form errors.

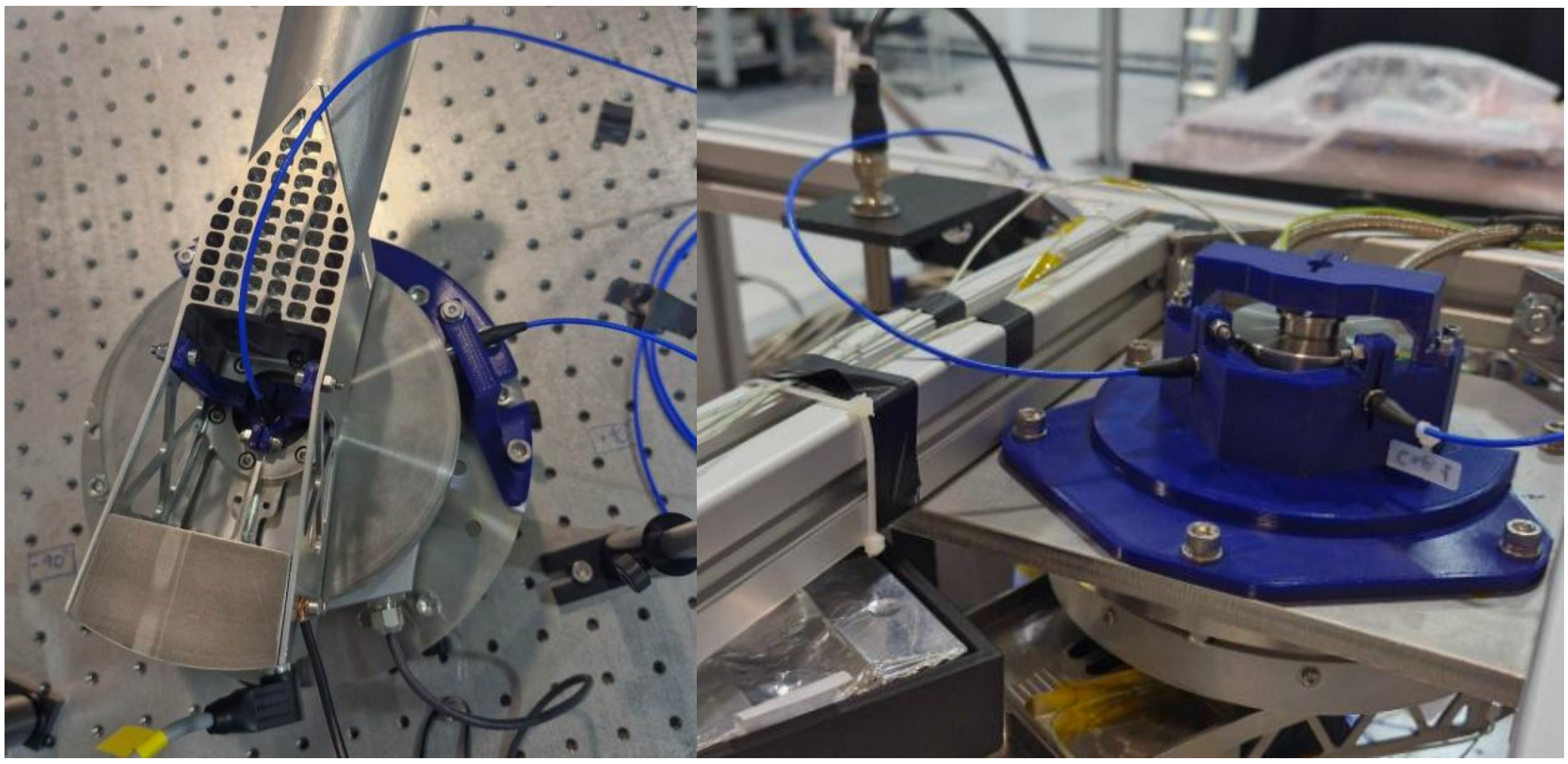

**Figure 5.** Radial runout experimental setup — capacitive sensors positioned 90° apart on the lateral surface of the shaft. Elbow (LPOA1E, left) and Shoulder (LPOA1S, right).

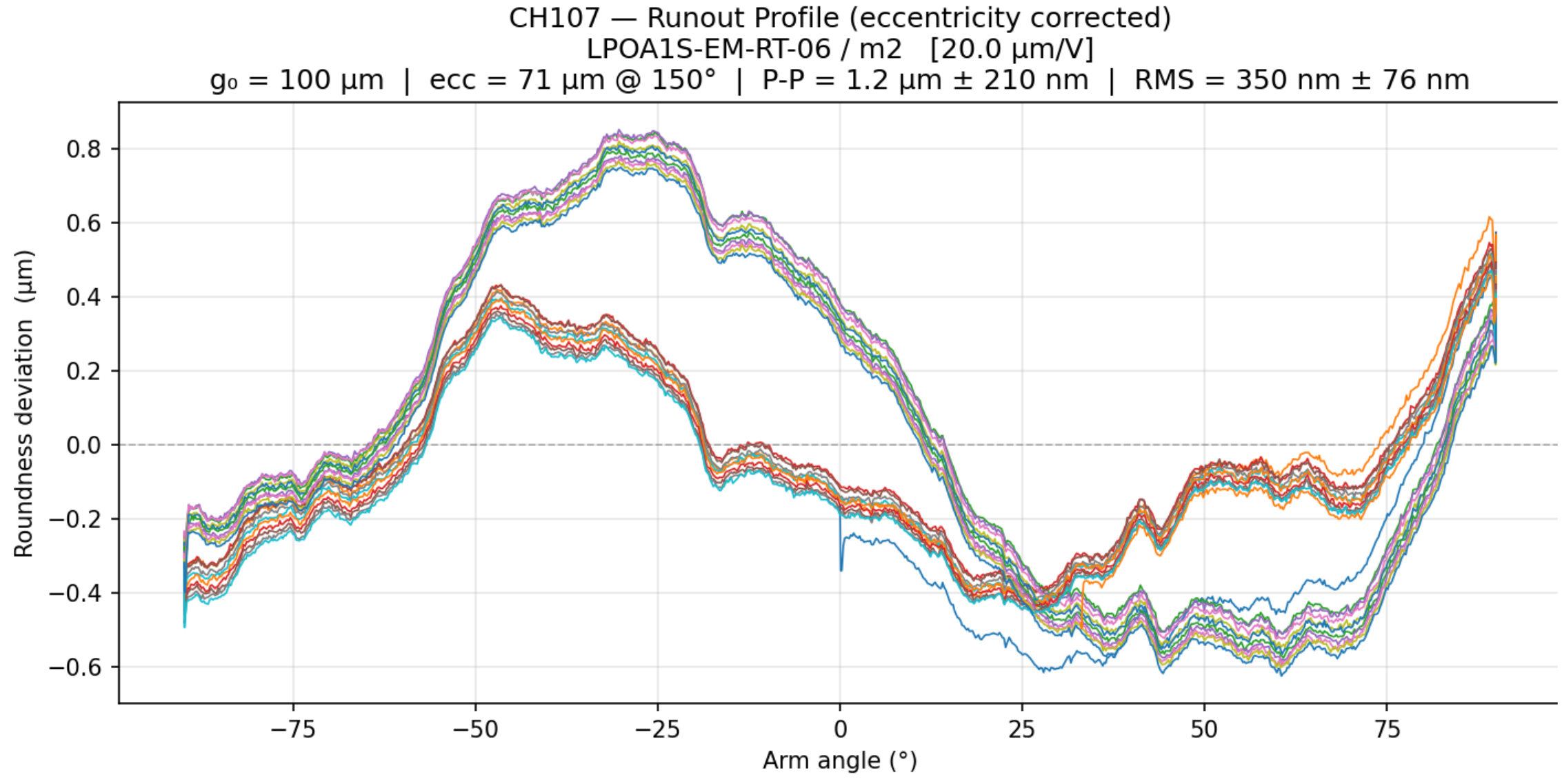

**Figure 6.** LPOAS Radial runout

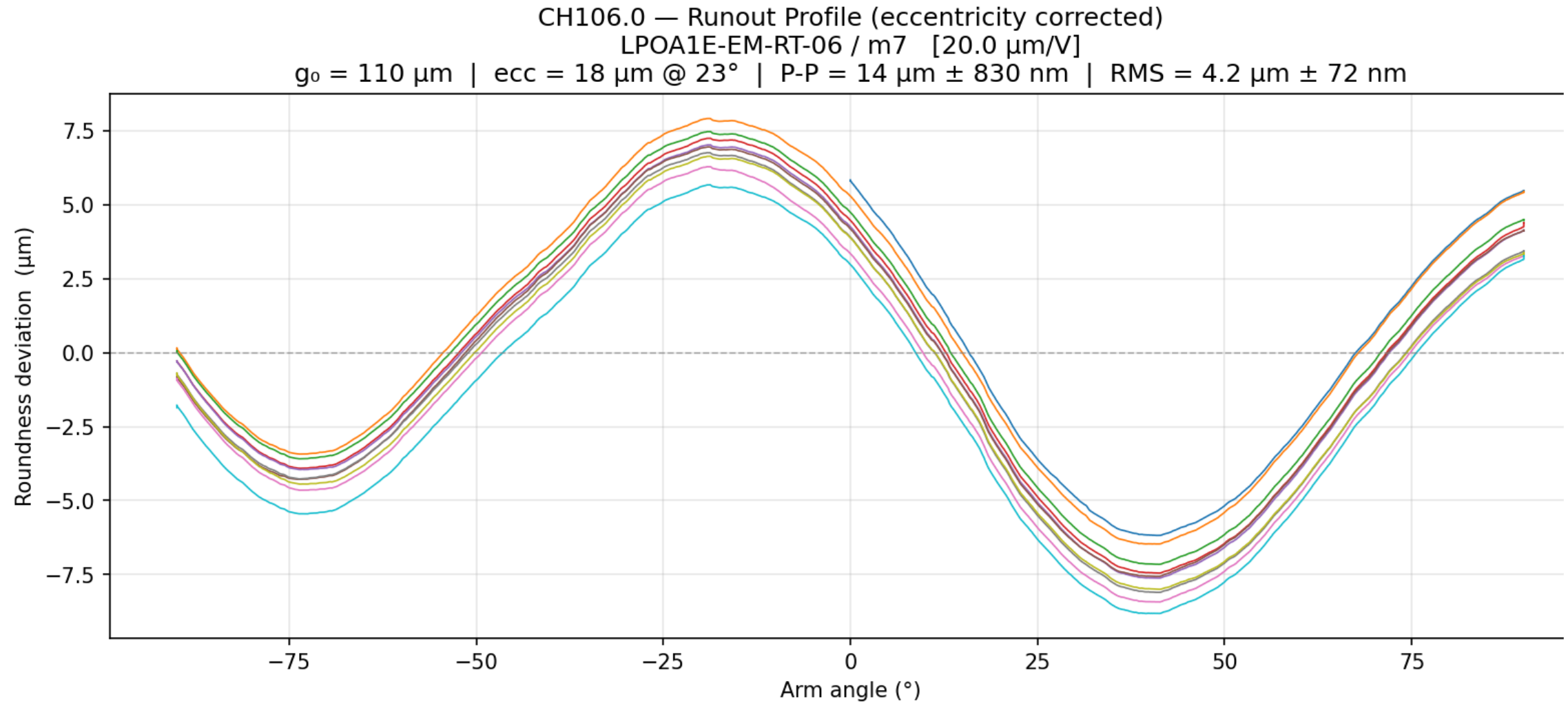

**Figure 7**. LPOAE Radial runout

### 3.4 Axial Runout

The axial runout was measured by directing the capacitive sensors towards the end face of the shaft during rotation (Fig 8). The signal contains a tilt component arising from the angular misalignment between the end face and the rotation axis, which is separated from the genuine face form error. The corrected axial profiles for the shoulder and elbow joints are presented in Fig. 9 and Fig. 10 respectively. The elbow end face runs notably flatter than the shoulder: the tilt component is 0.34 µm on the elbow versus 1.0 µm on the shoulder, and the peak-to-peak axial runout is 3.3 µm and 1.0 µm respectively.

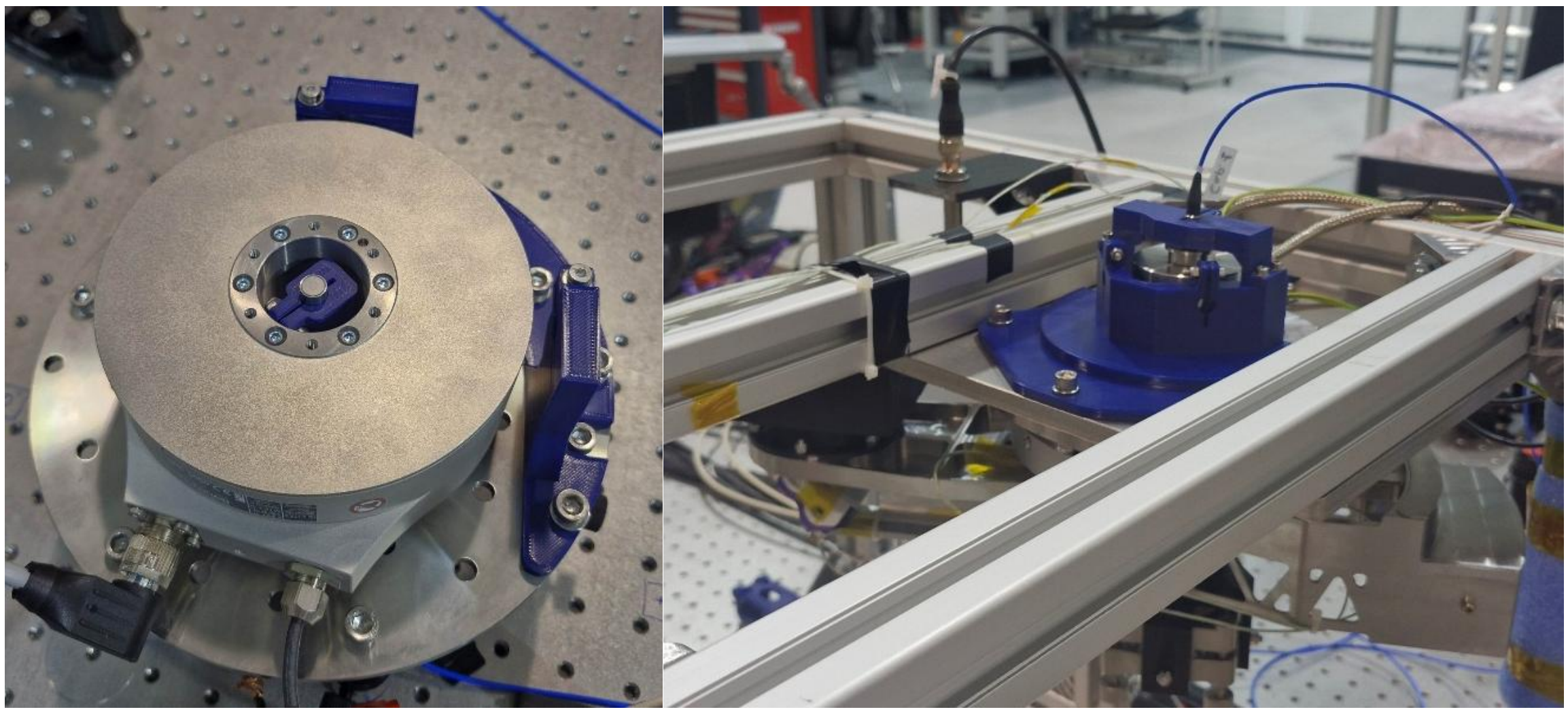

**Figure 8.** Axial runout experimental setup — capacitive sensors directed towards the end face of the shaft. Elbow (LPOA1E, left) and Shoulder (LPOA1S, right).

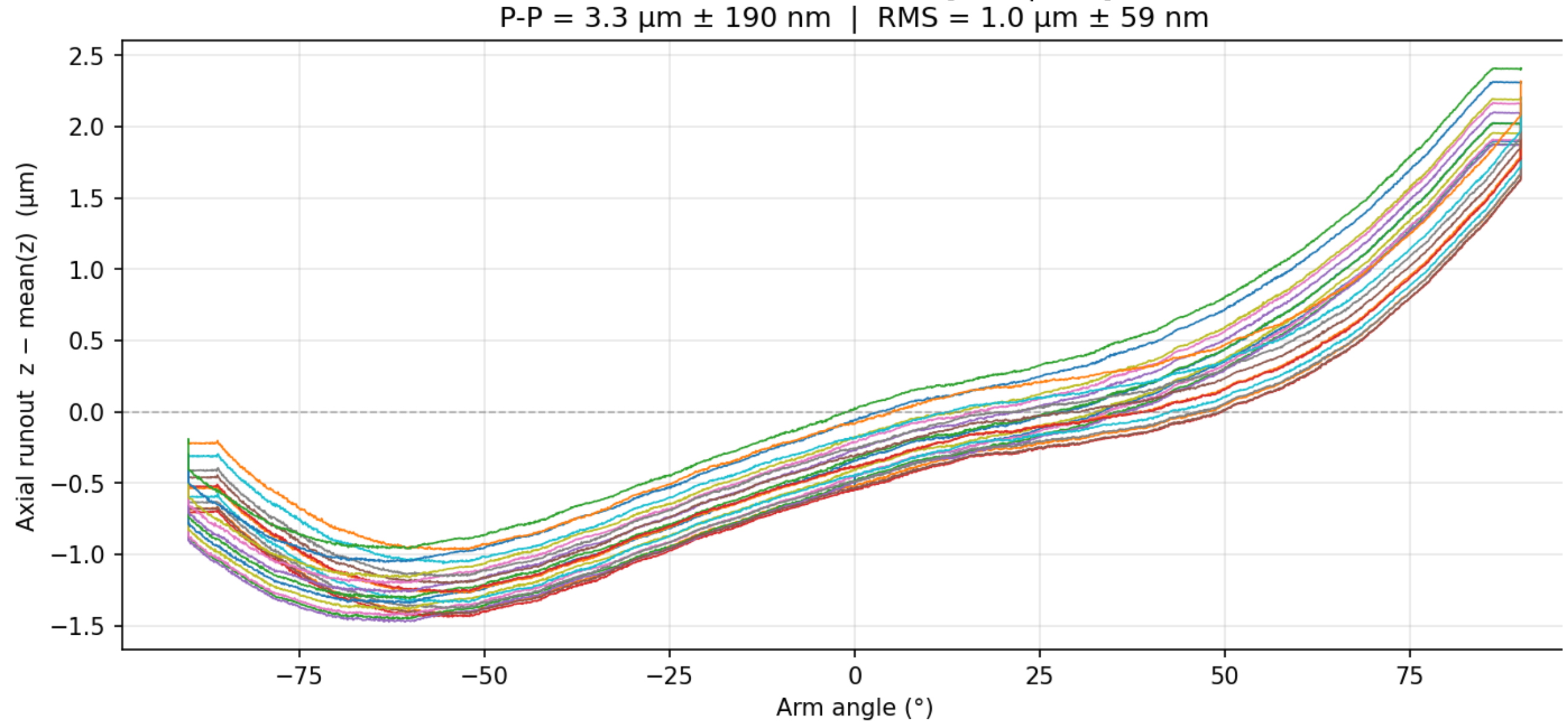

**Figure 9.** LPOAS Axial Runout

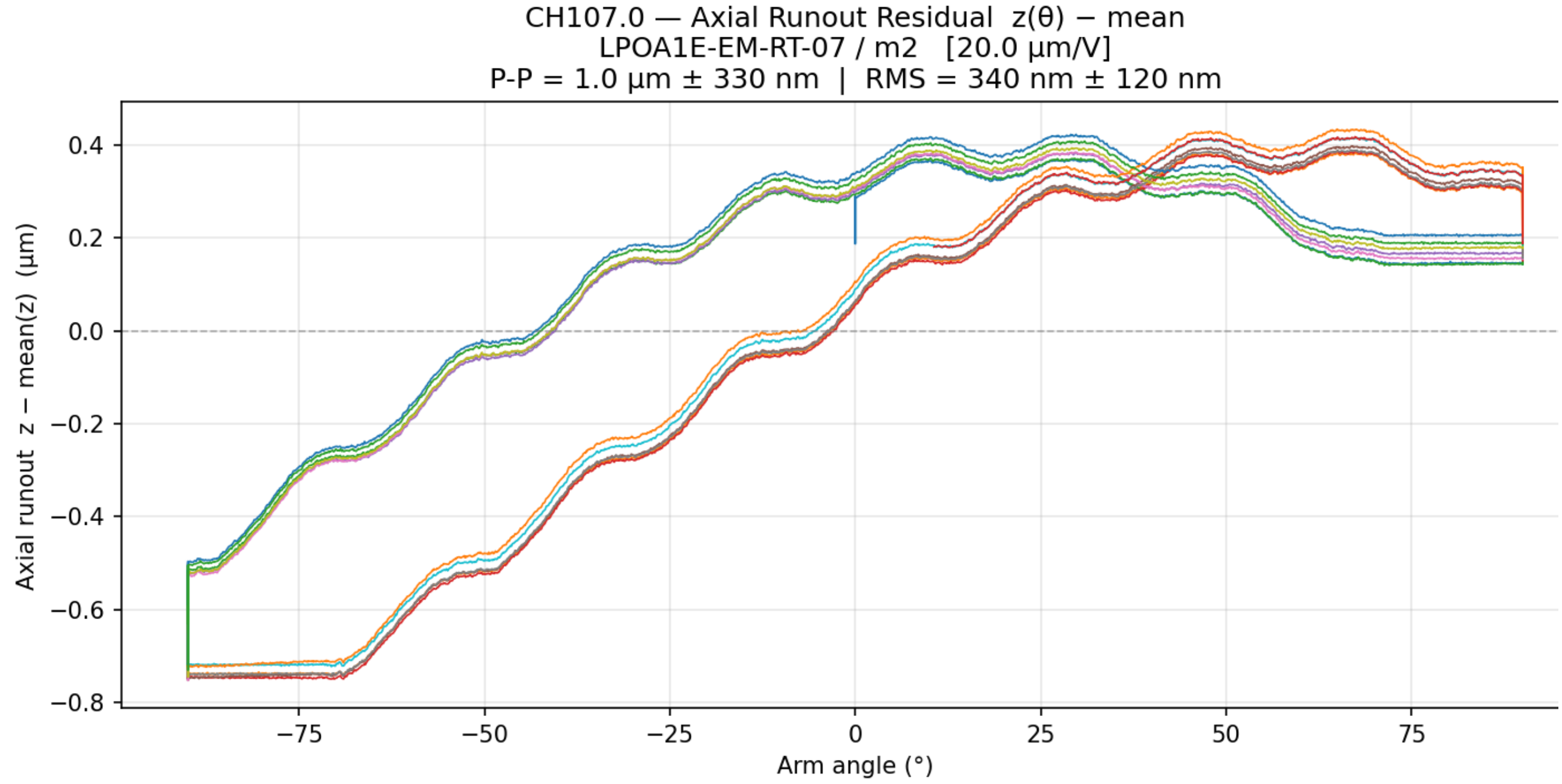

**Figure 10.** LPOAS Axial Runout

## 4. CONCLUSIONS

This proceeding has presented preliminary results from the mechanical characterisation campaign of the HARMONI LPOA Engineering Model, covering angular resolution, step tracking, rotation axis wobble, and radial and axial runout on the shoulder and elbow joints.

The angular resolution measurements show that both joints reliably resolve micro-steps well below the arcsecond level, with the shoulder achieving a noise-limited resolution of 0.08 µrad, approximately 25 times finer than the elbow. On the elbow joint, the encoder response was independently confirmed by an interferometer, demonstrating that the encoder faithfully reports the displacement delivered by the joint.

The wobble characterisation of the shoulder joint reveals that the rotation axis tilt is dominated by a single harmonic component, with higher-order residuals below 1.3% of the total amplitude. This indicates clean, repeatable bearing behaviour that is well-suited to harmonic error compensation in the control system.

Radial runout measurements show a repeatable surface form error of 1.2 µm peak-to-peak on the shoulder and 14 µm on the elbow. Axial runout is below 3.0 µm peak-to-peak on both joints, with the elbow end face running notably flatter than the shoulder.

Taken together, these results demonstrate that the selected design meets the precision requirements of the LPOA pointing mechanism. The campaign is ongoing; wobble measurements on the elbow joint and additional cross-validation with both instruments are planned as next steps.

## ACKNOWLEDGEMENTS

This work has been supported by grants PID2022-140483NB-C22 and PID2024-158231NB-C22, funded by AEI 10.13039/501100011033 and the European Union, and by grant BDC20221289, funded by the Spanish Ministry of Science and Innovation through the Recovery, Transformation and Resilience Plan, and by NextGenerationEU from the European Union through the Recovery and Resilience Facility